# IS SOCIAL MEDIA HINDERING OR HELPING ACADEMIC PERFORMANCE? A CASE STUDY OF WALTER SISULU UNIVERSITY BUFFALO CITY CAMPUS

Jose M. Lukose and Abayomi O. Agbeyangi
Walter Sisulu University, South Africa
Corresponding: aagbeyangi@wsu.ac.za

**ABSTRACT**

Social media platforms are popular among higher education students and have seen increased usage for academic purposes, especially during the COVID-19 pandemic. However, excessive use of social media can negatively impact students' academic performance. This preliminary study examines social media's impact on students' academic performance at Walter Sisulu University (WSU), Buffalo City campus. Using a positivist paradigm and a quantitative approach, randomly sampled data were collected from 71 students through a survey to identify trends and generate preliminary insights. Results indicate that while social media can facilitate academic work, it predominantly acts as a distraction, negatively affecting academic performance, particularly for first-year students. Notably, 84.5% of the students spend more than four hours daily on social media, and 39.4% agree that it negatively impacts their assignment completion. The study underscores the need for students to balance their social media use and academic responsibilities, highlighting the importance of this issue. Recommendations for achieving this balance, such as adopting time management strategies and integrating social media into teaching methodologies, are discussed.

**INTRODUCTION**

Social media, a game-changer in communication and information exchange, has become integral to daily life, including educational contexts (Boyd and Ellison, 2007; Odinokaya et al., 2022). The significant increase in social media use for educational purposes during the COVID-19 pandemic (Tkacová et al., 2022) has opened up new possibilities. Students and teachers can now stay connected, share materials, and collaborate on projects through social media. Tkacová et al. (2022) argue that credible social media platforms have the potential to transform students from passive participants to active collaborators, inspiring a new way of learning. Similarly, the widespread use of social media in other human activities, including education, has been observed. In most places worldwide, youths spend more time on social media than ever (Neelakandan et al., 2020; Odgers and Jensen, 2020; Adorjan and Ricciardelli, 2021). Platforms like Facebook, TikTok, X (formerly Twitter), Instagram, and Snapchat have become extremely popular among young people (Adorjan and Ricciardelli, 2021). Sharing short and viral videos, skits, and memes on these platforms further contributes to social media's engaging and entertaining nature, making them more appealing to the audience.

In today's world, academic achievements can help predict opportunities and success (Biggs, Tang, and Kennedy, 2022). At a pace never seen before, technological innovation is transforming every aspect of professions and our daily lives. As a result, preparing young people for a fast-changing world necessitates high-quality education free from social distractions. Those who lack a solid academic background may find it challenging to participate in the highly competitive economy and adjust to the often shifting needs of work (Neelakandan et al., 2020). By developing a culture of concentration and commitment, students can better prepare for success in a world that is becoming increasingly digital and fast-paced. However, social media's potential to distract students is significant. Excessive use can lead to addiction, reduced academic performance, and neglect of essential activities like assignments and real-life interactions (Amin et al. 2016). During the COVID-19 pandemic, most universities promoted social media for collaboration and communication, but the effect of this increased usage on academic performance is uncertain. Both educators and parents continue to express concerns that students may not be using social media for educational purposes or dedicating enough time to academic activity. Hence, educators and parents must actively guide students' use of social media, helping them understand its potential benefits and risks and encourage a balanced approach. The need for more research on this problem in South Africa highlights the urgency of investigating the influence of social media on academic performance.

This preliminary study aims to investigate the effects of social media usage on students' academic performance at Walter Sisulu University's (WSU) Buffalo City campus, providing insights into its positive and negative impacts. Walter Sisulu University, located in the heart of East London, South Africa, is known for its diverse student body and commitment to integrating technology into education. The Buffalo City campus, in particular, offers a vibrant academic environment that strongly emphasises utilising digital tools to enhance learning. This makes it an ideal location for studying social media's dual role in facilitating academic engagement and potentially serving as a distraction. Focusing on this campus aims to generate relevant insights that can benefit similar regional institutions. The study examines the balance between social media's positive and negative impacts by evaluating students' usage patterns and their effects on academic performance. Notably, platforms such as Facebook, TikTok, and Instagram are identified as particularly prevalent and problematic among WSU students.

The paper's other specific contributions are to (1) provide insights that can inform interventions and policies aimed at maximising the positive impacts of social media. (2) examine the relationship between different types of social media use and academic performance. Specifically focusing on the research questions:
- What is the level of exposure to social media platforms among students?
- What do the students use social media for?
- How does social media usage influence the academic performance of students?

The remaining parts of the paper are as follows: section 2 discusses the theoretical background and related work; Section 3 details the methodology; Section 4 discusses the results; and Section 5 concludes the paper.

**BACKGROUND AND RELATED WORK**

**Theoretical Background**

The study explores the connection between student academic performance and social media use. Social Learning Theory (SLT) (Bandura and Walters 1977) and Uses and Gratifications Theory (Katz, Blumler, and Gurevitch 1973) provide the theoretical foundation for this study to understand the interactions between academic performance and learning habits. SLT emphasises the significance of modelling and observation in influencing behaviour, whereas

UGT underscores the active role of individuals in choosing and utilising media to satisfy their needs.

SLT stresses the need for learning through imitation and observation. Students can expose themselves to a variety of academic activities through social media. They can see classmates succeed academically, use their time well, or even put things off. These observed behaviours, as models, may encourage students to follow suit; the same is true for negative behaviours. Bandura (2016) emphasises that individuals can learn through direct experience and by observing and imitating others, stressing the impact of the four core processes: attentional processes, representational processes, translational production processes, and motivational processes. For instance, a friend's successful use of online flashcards might encourage a student to attempt them. Conversely, seeing others use social media could make procrastination acceptable or cause feelings of slipping behind. However, SLT makes no assumptions about passive learning. According to the theory, students choose the models they see as relevant or successful based on their social media experiences (Rothkrantz 2015).

On the other hand, UGT concentrates on the user's point of view (Ruggiero 2000; Rrustemi et al. 2021). It implies that people deliberately select media outlets to satisfy specific needs and wants. According to Rrustemi et al. (2021), individuals' deliberate selection of media outlets is driven by their desire to fulfil specific needs and wants, whether for entertainment, information, social connection, or other gratifications. Hence, students may use social media to pursue different kinds of satisfaction. Some might connect with classmates for cooperative learning or look for study materials. Others may use it for social interaction or entertainment, potentially leading to distractions. The theory suggests that media exposure gradually shapes individuals' perceptions of reality. Users tend to select media platforms aligning with their established beliefs and values, fostering specific attitudes and worldviews. In cognitive dissonance, the theory posits that individuals actively seek media platforms that mirror their beliefs and attitudes, avoiding information that challenges their worldview to maintain psychological harmony. The selective exposure theory further suggests that people actively seek media content that validates their beliefs and attitudes while shunning information that questions or contradicts their views. This behaviour is driven by a desire to have their chosen media platform reinforce their current perspectives and avoid exposure to dissenting opinions, thereby creating a self-reinforcing cycle.

This study achieved a more theoretical overview of how social media influences student learning by integrating the perspectives of Social Learning Theory (SLT) and Uses and Gratifications Theory (UGT). Social Learning Theory (SLT) and Uses and Gratifications Theory (UGT) inform our research questions by providing a framework to understand students' social media usage and its impact on academic performance. SLT helps us explore how students' exposure to social media (Research Question 1) influences their learning behaviours through observation and imitation, affecting their academic habits and performance (Research Question 3). UGT examines the motivations behind students' social media use (Research Question 2), revealing how their need for information, social interaction, or entertainment influences their academic engagement. Together, these theories help analyse social media's positive and negative effects on students' academic outcomes. For example, students with a solid drive to improve their grades (UGT) are more likely to actively seek out and learn from positive academic role models they encounter on social media (SLT).

**Related work**

The impact of social media on student academic performance has been a topic of growing interest in recent years. Some of these studies highlight the potential advantages of social media for student learning. Ansari and Khan (2020) explore how students can leverage social media platforms for collaborative learning activities. The study found that using social media in collaborative learning within higher education institutions significantly impacts students' engagement and academic performance. The study highlights the importance of utilising social media platforms to transfer course materials, facilitate collaborative learning, and enhance interaction among students and teachers. Similarly, Troussas et al. (2020) examine how students utilise *i-LearnC*, an intelligent tutoring application developed for learning programming over social media. The study demonstrates the potential for social networks to serve as effective platforms for intelligent and adaptive tutoring in higher education, promoting effective learning and teaching. Khan et al. (2021) investigate the effect of the COVID-19 pandemic on collaborative learning-driven social media adoption. The study findings suggest that social media usage positively influences factors such as perceived ease of use, social media anxiety, and perceived usefulness, ultimately enhancing the overall learning experience.

Social media usage for learning purposes has been found to correlate with academic performance positively. In their 2019 study, Gloria and Akbar (2019) examined the relationship between the amount of time spent on social media for learning or entertainment and academic performance, as indicated by the Grade Point Average (GPA) of 126 fourth-year undergraduate students. The results showed a noteworthy positive correlation between using social media for educational purposes and academic performance. This suggests that students who engaged more with educational content on social media tended to have higher academic achievements. In Goet's (2022) study, the impact of social media on students' academic performance is examined. The findings indicate that students frequently use social media to connect with peers, carry out research for their assignments, access educational materials, stay informed about current trends, and typically spend 2 to 4 hours on these platforms each day. The study underscores the importance of schools establishing guidelines for student social media use to promote social media literacy in conjunction with digital literacy education and foster responsible and ethical social media and technology usage. Also, Al-Menayes (2015) and Lau (2017) contended that social media exerts a profound influence on young adults, offering educational advantages through the facilitation of enhanced communication, the provision of networking opportunities, and the expedient dissemination of resources. These findings suggest that while social media can be a valuable tool for enhancing learning, its impact on academic performance depends on how it is used.

However, research also identifies potential drawbacks associated with social media use. Studies like Tkacová et al. (2022) report a negative relation between excessive social media use and academic performance. This negative association might be attributed to increased distractions or procrastination behaviours triggered by social media engagement. It is evident from the research that excessive social media use can indeed have an adverse impact on academic performance, highlighting the need for moderation and balance in its usage among students. Adorjan and Ricciardelli (2021) support this notion, highlighting how the constant notifications and readily available entertainment on social media platforms can lead to students spending less time on academic tasks and experiencing decreased focus. Similarly, Griffioen et al. (2021)further assert that the habit-forming nature of social media use, often driven by boredom or habit, can draw students away from focusing on academic responsibilities. It emphasises that variability in social media usage patterns, including the time spent and number of platforms used, suggests that students may allocate more time to social media than academic pursuits.

While there is a growing body of research on social media and academic performance globally, studies specifically focused on the South African context are somewhat limited.

Bamigboye and Olusesan (2017) investigate the impact of social media on learning in Eastern Cape Universities, South Africa. The study found no significant difference in students' acceptance and impact of social media on learning, as it offers a blend of traditional and modern learning experiences. Similarly, Lottering (2020) demonstrates that social media platforms offer significant advantages in enhancing student engagement with course material and lecturers, leading to various educational benefits like increased student interest, improved academic performance, and enhanced critical thinking skills. Consequently, social media use patterns, educational environments, and cultural factors can vary significantly across countries. Additionally, Lukose et al.'s (2023) study on the impact of social media on young adults in Buffalo City, South Africa, found a link between prolonged social media use and negative mental health outcomes like depression, anxiety and suicidal thoughts. Recommendations include awareness campaigns to educate young adults on the risks of overusing social media and promote healthier online behaviours to safeguard mental well-being. Notably, in the South African context, particularly focusing on tertiary education, key factors include the diverse student demographics, varying levels of access to digital resources, and the specific challenges South African students face, such as socio-economic disparities and limited access to high-speed internet. These factors significantly shape the use of social media, a tool that profoundly influences academic performance in South Africa, making it an urgent issue to address.

The existing research offers valuable insights into the connection between social media and academic performance. While social media can enhance learning, its potential for distraction and negative influences cannot be ignored. The limited research within the South African context highlights the need for further investigation to understand how these dynamics play out in this specific educational landscape. Additionally, there are limited studies that simultaneously consider both the positive and negative impacts of social media, which our research addresses by providing a balanced view. This study aims to contribute to these gaps in knowledge and offers valuable guidance for promoting appropriate social media use among South African students to optimise their academic success.

The related studies align closely with our investigation by highlighting the dual role of social media in both enhancing and hindering student academic performance, which is a core focus of our research at WSU. Studies by Ansari and Khan (2020) and Troussas et al. (2020) demonstrate social media's potential as a collaborative learning tool, boosting engagement and academic outcomes, similar to our findings. Conversely, research by Tkacová et al. (2022) and Adorjan and Ricciardelli (2021) underscore the risks of excessive social media use, such as distractions and reduced academic focus, which we also observed. Furthermore, studies like Bamigboye and Olusesan (2017) and Lottering (2020) contextualise the unique challenges in the South African educational landscape, emphasising socio-economic disparities and varying digital resource access, factors that our research specifically addresses in examining the impact of social media on WSU students. Collectively, these studies reinforce our comprehensive approach to understanding the nuanced effects of social media on academic performance, particularly within the South African context.

**METHODOLOGY**

The study used a quantitative approach as a case study, aiming to objectively measure the impact of social media on academic performance using numerical data and statistical analysis. A survey design was chosen, using random sampling to collect data from a sample of 71 students at Walter Sisulu University's Buffalo City campus.

**Research Instrument**

Based on the study's research objectives, the researchers designed a self-administered questionnaire to collect data. The questionnaire consists of five different sections. The first section gathered socio-demographic data, including gender, age, and educational level. The remaining sections focus on social media types, devices used, impact, and relationship to academic performance.

**Data Collection**

The data from the respondents were collected using a survey that included a self-administered questionnaire. The questionnaire featured closed-ended questions to ensure uniformity in responses and ease of analysis. Participants were asked about their social media usage, including the types of platforms used, duration of use, and purposes (e.g., socialising, academic, professional). Additionally, they were asked about their academic performance and how they perceived social media to affect their studies. Given the exploratory nature of this research, the sample size is considered sufficient for identifying trends and generating preliminary insights (Acharya et al. 2013).

The participants received sufficient information about the study's objectives. Consistent with research ethics, each participant's consent was obtained verbally. Participants were guaranteed confidentiality, anonymity, and privacy in their profiles and responses. The study followed ethical guidelines when conducting the data collection.

**Data Analysis**

Descriptive and inferential statistical analysis was conducted using both Microsoft Excel and SPSS. Descriptive statistics were employed to ascertain the link between the variables and to analyse the trends. The data collection instrument's reliability and validity were tested using SPSS.

**Study Design**

The study's exploratory nature allows for the examination and explanation of links between the identified variables, specifically their cause-and-effect relationships. The design is described as follows:

Let $N$ represent the total population of students at the Walter Sisulu University Buffalo City campus. A sample $n$ of 71 students was selected and given as:

$$n = 71$$

Each student's response was recorded for various variables, including:

- Social Media Usage Frequency ($F_i$): The number of hours a student $i$ spends on social media per day.
- Social Media Platforms ($S_i$): The specific social media platforms used by students $i$.
- Purpose of Social Media Use ($P_i$): The reasons for social media use by students $i$ (e.g., socialising, academic purposes, entertainment).

- Academic Performance ($G_i$): Measured by self-reported grades or GPA of student $i$.

The data analysis is focused on answering the following research questions:

A. **Level of Exposure Towards Social Media Platforms:**
   - Frequency and Duration: How frequently and for how long do students use social media?

$$E_i = F_i$$

$E_i$ represents the level of social media exposure for student $i$.

$$Mean\ Exposure = \frac{1}{n}\sum_{i=1}^{n} E_i$$

   - Platform Popularity: Which social media platforms are most popular among students?

$$S_i = Set\ of\ platforms\ used\ by\ student\ i$$
$$Platform\ Popularity = \{S_i\}\ for\ i = 1, 2, \dots, n$$
$$Count\ of\ Each\ Platform = Count(S_i)$$

B. **Purpose of Social Media Use:**
   - Purpose Categories: The primary purposes for which students use social media are categorised as socialising, academic purposes, or entertainment.

$$P_i = \{P_{i1}, P_{i2}, \dots, P_{im}\}$$
$$Purpose\ Distribution = \{P_i\}\ for\ i = 1, 2, \dots, n$$
$$Count\ of\ Each\ Purpose = Count(P_{ij})$$

Where $P_{ij}$ represents the purpose $j$ (e.g., socialising, academic purposes) for student $i$.

C. **Influence on Academic Performance:**
   - Perceived Impact: How students perceive the impact of social media usage on their academic performance.

$$A_i = G_i$$

$$Mean\ Academic\ Performance = \frac{1}{n}\sum_{i=1}^{n} A_i$$

The study design focuses on the three primary research questions: the level of exposure to social media platforms among students ($E_i = F_i$), the purposes for which students use social media ($P_i = \{P_{i1}, P_{i2}, \dots, P_{im}\}$), and how social media usage influences academic performance ($A_i = G_i$). Descriptive analysis is utilised to summarise the data and identify trends, offering valuable insights into the relationship between social media and academic performance.

## RESULTS AND DISCUSSION

### Demographic Analysis

The demographic analysis of the respondents, as shown in Table 1, includes age, gender, and academic level. The age distribution indicates a majority of respondents aged 20-24 (40.2%), followed by those aged 25-29 (22.9%), below 20 (19.1%), 30-34 (10.2%), and 35 or above (7.6%). The gender distribution suggests that 53.5% of respondents are male, while 46.5% are female. Regarding academic level, the majority of respondents are at Level 3 (53.5%), followed by Level 1 (25.4%), Level 2 (14.1%), and Level 4 (7%).

**Table 1:** Demographic Distribution of Respondents

| Socio-demographic data | | Count/Percentage (%) |
|---|---|---|
| **Age** | <20 | 19.1 |
| | 20-24 | 40.2 |
| | 25-29 | 22.9 |
| | 30-34 | 10.2 |
| | >35 | 7.6 |
| **Gender** | Male | 53.5 |
| | Female | 46.5 |
| **Academic Level** | Level-1 (undergraduate) | 25.4 |
| | Level-2 (undergraduate) | 14.1 |
| | Level-3 (undergraduate) | 53.5 |
| | Level-4 (Postgraduate) | 7.0 |

### Level of Exposure to Social Media Platforms

This analysis, as contained in Table 2, measures the degree of social media exposure among the students, focusing on the time spent on social media and the type of social media platforms they predominantly use. The results reveal that most respondents (97.2%) own smartphones, indicating easy access to social media, while only 2.8% do not own smartphones. Furthermore, 94.4% of respondents have at least one social media account, compared to 5.6% who do not own any social media accounts. In addition, from the choice of devices, most of the respondents selected smartphones as the most preferred device to access social media (77.1%), followed by laptops (67.1%) and tablets (41.4%). Despite this, the respondents frequently utilise various devices to access social media.

**Table 2:** Device and Access to Social Media Platforms

| Exposure Towards Social Media Platforms | | Count/Percentage (%) |
|---|---|---|
| **Smartphone** | Yes | 97.2 |
| | No | 2.8 |
| **Social Media Account** | Yes | 94.4 |
| | No | 5.6 |
| **Device use for social media** | Smartphone | 77.1 |
| | Computer | 67.1 |
| | Tablet | 41.4 |

**Analysis of the average frequency and duration of social media use**

This analysis examines respondents' average frequency and duration of social media use. A significant majority (84.5%) of the students spend more than 4 hours per day on social media, indicating high social media engagement. Regarding the perceived addiction to social media, the results show that 28 respondents (39.4%) agree, and 17 respondents (23.9%) strongly agree that they are addicted to social media. Meanwhile, 10 respondents (14.1%) are neutral, and a smaller portion either strongly disagree (11.3%) or disagree (11.3%) with the statement of being addicted. Figure 3 illustrates the analysis of the frequency and time spent on social media platforms among the respondents.

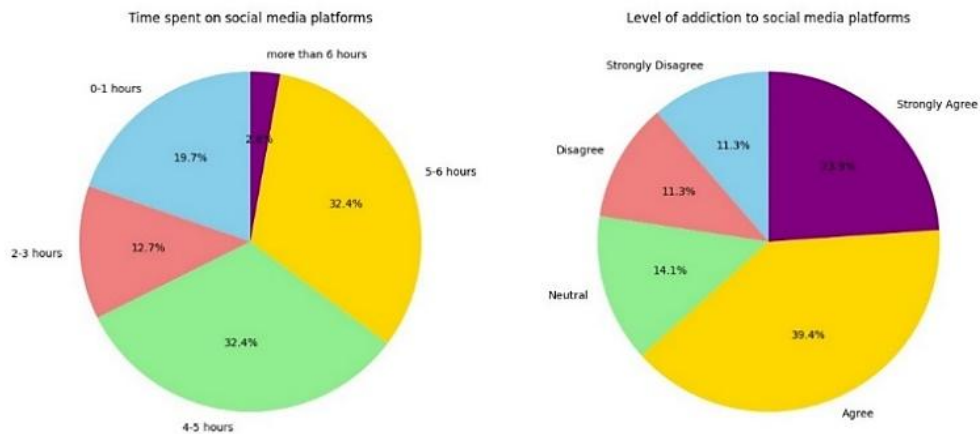

**Figure 3:** The average frequency and duration of social media use

**Identification of the most popular social media platforms among students.**

The analysis in Figure 4 indicates that WhatsApp and Facebook are the predominant social media sites among students, with a substantial majority utilising two or more channels. Figure 4(a) illustrates the distribution, with WhatsApp (88.7%) and Facebook (77.5%) being the most dominant platforms. Figure 4(b) examines why students use social media, indicating that the overwhelming majority (93%) use it for socialising, followed by academics (73.2%). This implies that social media substantially impacts the personal and academic aspects of students' lives.

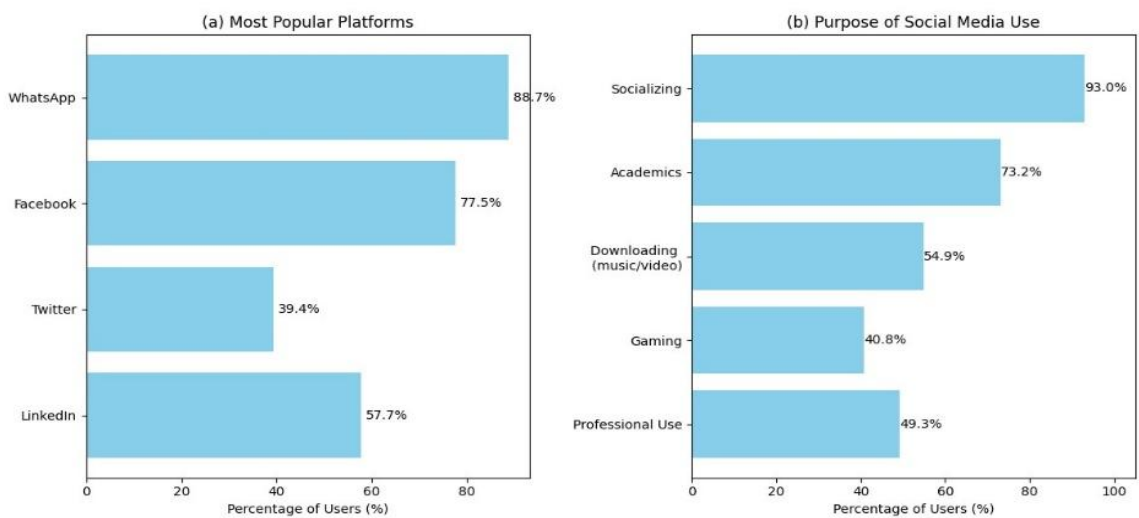

**Figure 4:** Popular Social Media Platforms and Usage

**Categorisation and analysis of the primary purposes for which students use social media**

Figure 5 reveals how students perceive social media for academic purposes, communication with lecturers, and collaborative work with peers. Figure 5(a) suggests a significant number (67.6%) find social media valuable for academic purposes, while a smaller percentage (15.5%) indicates some resistance. Similarly, the results on the usefulness of social media to communication with lecturers (62%) and collaborative work (76%) reveal a growing acceptance of social media in these contexts. Analysing these percentages provides valuable insights into how students view social media's role in their academic pursuits.

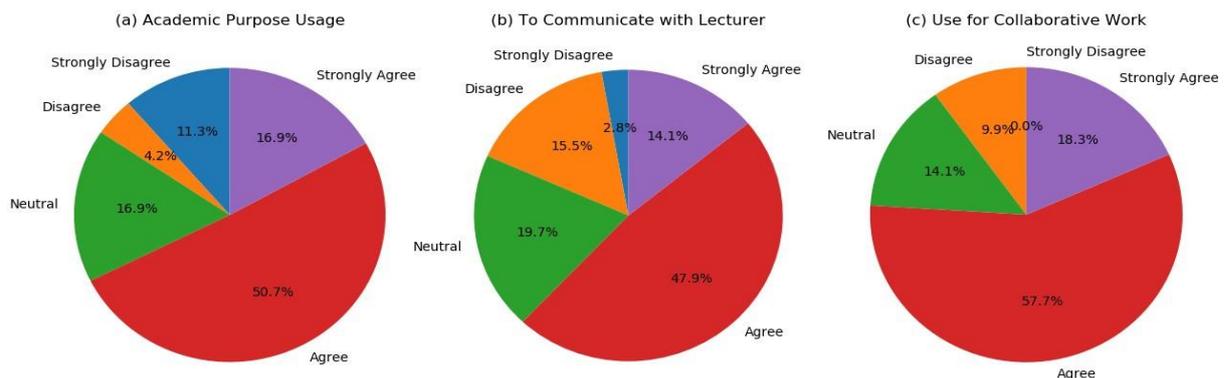

**Figure 5:** Primary Purpose of Usage

**Examination of the perceived impact of social media on students' academic performance.**

The results highlight social media's significant impact on students' academic performance. A majority of the respondents (57.7%) believe that their understanding of academic topics improves through social media discussions with peers and lecturers, while a minority (19.7%) disagree. Furthermore, 62% of the respondents are open to increasing the use of social media in classrooms, with only 21.2% opposing this idea, indicating some acceptance of integrating social media into classroom settings. Additionally, 77.4% of respondents find academic materials obtained from social media useful. In contrast, only 8.5% oppose the idea, suggesting that most students value the additional resources and perspectives offered by social media platforms. These findings underscore the potential of social media as a valuable tool for enhancing learning and engagement among students.

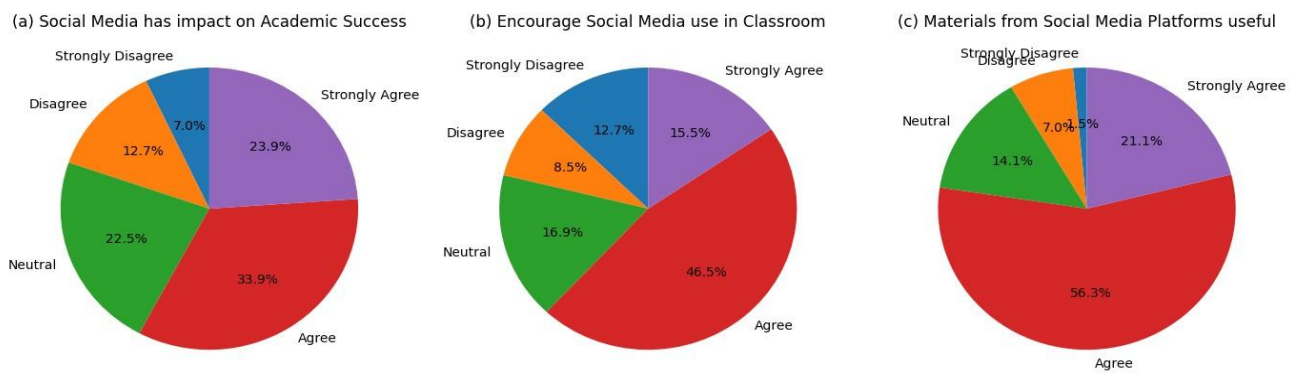

**Figure 6**: Perceived impact of social media on students' academic performance

**Analysis of any observable trends between social media usage and academic performance.**

The results in Table 1 reveal a strong correlation between social media use and potential negative impacts on academic performance. A substantial majority (64.8%) of the respondents agreed that social media leads to procrastination, while a small number (8.4%) disagreed. This suggests that social media use might be a significant factor contributing to procrastination. Furthermore, 70.5% of the respondents agree that sharing lecture materials on social media leads to decreased lecture attendance, while only 15.5% disagree. Also, in the Table, the impact of social media on assignment turnout by students is measured, and most of the respondents (71.8%) agree that social media can have a negative effect on assignment turnout. The effect of time spent on social media and academic work shows that the respondents (64.3%) agree they spend most of their time on social media rather than academics, with only 11.2% disagreeing.

**Table 3:** Observable Trends Between Social Media Usage and Academic Performance

| Statement | Strongly Disagree | Disagree | Neutral | Agree | Strongly Agree |
|---|---|---|---|---|---|
| (a) Social media can lead to Procrastination | 5% | 3.4% | 26.8% | 39.4% | 25.4% |
| (b) Sharing lecture material on social media can reduce lecture attendance | 8.5% | 7% | 14.1% | 42.3% | 28.2% |
| (c) Excessive use of social media can affect assignment | 4.3% | 7% | 16.9% | 49.3% | 22.5% |
| (d) Time spent on social media can affect academic work | 2.8% | 8.6% | 24.3% | 40% | 24.3% |

**Analysis of Perceived General Impact of Social Media**

The analysis is to determine whether there has been any improvement in academic performance since the respondents became acquainted with social media. Specifically, Figure 8(a) depicts the diverse degrees of influence that social media usage has on academic achievement among different student categories, ranging from freshmen to seniors. It suggests that a considerable proportion of students view social media as having a significant influence on their academic achievement. More precisely, 30% of first-year students and 25% of second-year students indicate a moderately significant influence, whereas 25% of first-year students and 15% of second-year students indicate an extremely significant influence. Nevertheless, this pattern decreases marginally, with 20% of third-year and 15% of postgraduate students indicating a moderately significant influence, and only 10% of third-year and 5% of postgraduate students experience an extremely substantial impact. This implies that older students may either adjust more effectively to control their social media usage or have a greater understanding of its potential distractions, thereby decreasing its perceived influence on their academic achievements. Figure 8(b) examines students' perspectives on the impact of social media on their capacity to do academic tasks, such as assignments. The data provides a more diverse viewpoint in contrast to Figure 8(a). Approximately 20% of first-year and 25% of second-year students believe that social media has a somewhat negative impact on their capacity to accomplish assignments effectively. Among these, 5% of first-year and 10% of second-year students indicate a substantial decline in their productivity. Conversely, a significant change was observed among third-year and postgraduate students. Specifically, 25% of third-year and 15% of postgraduates indicate a moderate improvement in their task completion ability, while 20% of third-year and 15% of postgraduates claim a substantial improvement. This suggests a potential adjustment or deliberate use of social media for educational objectives in the long run. These observations emphasise the contrasting characteristics of social media: it has the potential to impede the completion of academic tasks for inexperienced students, but it may also be a valuable tool for those who have successfully incorporated it into their academic routines.

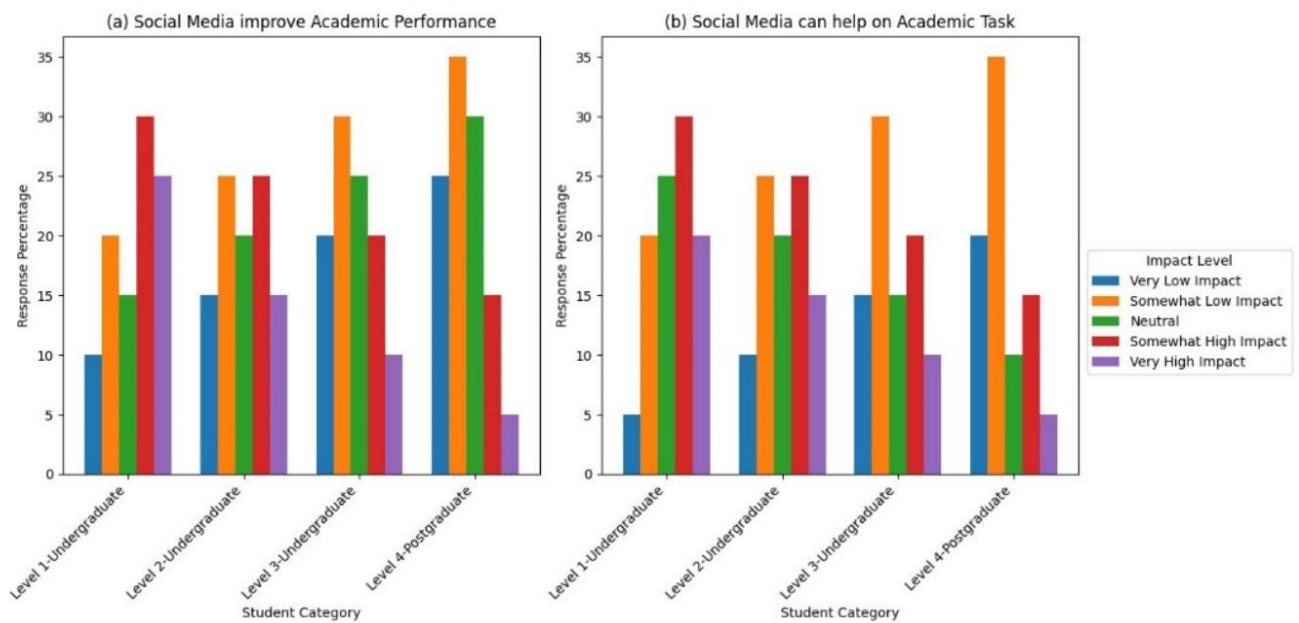

**Figure 8:** Perceived Positive Impact of Social Media on Academic Performance

**DISCUSSION**

**What is the level of exposure towards social media platforms among students?**

The study reveals significant engagement with social media platforms among the randomly selected students. Almost all participants (97.2%) possess smartphones, which serve as their primary device for accessing social media. A significant majority (94.4%) of the students have at least one social media account, indicating their active involvement with these platforms. This suggests a highly interconnected social media presence among the student body.

In terms of utilisation, the results indicate a focus on social interaction and academic activities. WhatsApp and Facebook are the primary platforms, with a significant percentage of students using multiple platforms. A majority of students (93%) use social media primarily for socialising, whereas a slightly lower percentage (73.2%) use it for academic purposes compared to socialising. This implies a substantial influence on both the personal and academic components of the student's lifestyle.

**What do the students use social media for?**

The study findings show a multifaceted approach to social media usage among students. While social interaction remains the primary purpose, with WhatsApp and Facebook being the most popular platforms, a significant portion (73.2%) also use social media for academic purposes. This suggests that social media has become an important tool for both personal connections and academic pursuits in student life.

The results present a complex picture regarding the influence of social media on academic performance. A majority (57.7%) believe social media discussions enhance their understanding of academic topics, and most students (77.4%) find materials obtained through these platforms useful. Additionally, a substantial number (62%) are open to increased social media use in classrooms, suggesting potential for enhanced learning and engagement. On the other hand, the research highlights potential drawbacks. A large proportion of respondents (64.8%) believe social media leads to procrastination, and an even greater percentage (70.5%) agree that sharing lecture materials online reduces attendance. Furthermore, the majority (71.8%) perceive social media's negative impact on assignment completion, and a significant number (64.3%) feel they spend more time on social media than academics.

**How does social media usage influence students' academic performance?**

The study reveals a challenging perspective on the impact of social media on academic achievement. There are both advantages and disadvantages. The majority of students believe that social media has the potential to improve their academic experience. 57.7% of students perceive advantages in engaging in conversations with their peers and lecturers. Additionally, 77.4% of students recognise the value of educational resources acquired via social media. Furthermore, 62% of students are receptive to the idea of including more social media usage in the classroom. These findings indicate the possibility of enhancing learning and involvement. However, the research also underscores possible adverse effects. Approximately two-thirds of the participants link social media usage with procrastination (64.8%), and a significant majority (70.5%) report decreased attendance in lectures as a result of accessing shared content through social media. Also, 71.8% of students perceive a negative impact on assignment completion, and 64.3% perceive a negative impact on their academic time. These findings indicate that social media can serve as a source of diversion and impede concentration on academic tasks.

In general, the study further suggests a potential learning curve. Level 1 (first-year) and 2 (second-year) undergraduate students seem more susceptible to the negative impacts, while level 3 (third-year) undergraduate and postgraduate students report either improved task completion skills or a lesser negative impact. This could indicate that students develop strategies to manage their social media usage over time or learn to use it for educational purposes. More so, social media plays a significant role in student life, offering opportunities for connection and learning alongside potential distractions. Understanding these complexities can help students develop strategies to leverage the benefits of social media while mitigating its drawbacks to optimise their academic performance. It is evident that the impact of social media usage on academic performance varies across different demographic groups. Younger students, particularly those below 20 and in the 20-24 age range, tend to use social media more frequently, which can provide learning opportunities and act as a source of distraction. Gender differences in social media usage are minimal, but preferences for platforms and usage purposes may vary, influencing academic outcomes. Students at higher academic levels (Levels 3 and 4) often utilise social media for academic collaboration, while those at lower levels (Levels 1 and 2) may use it more for social interaction. These findings can inform interventions aimed at maximising the benefits and mitigating the negative effects of social media on academic performance.

**Comparison with existing studies**

Our study's findings on the impact of social media on academic performance at Walter Sisulu University (WSU) align with previous research while also contributing new insights, particularly within the South African context. Ansari and Khan (2020) identified the dual role of social media as both a facilitator and a distraction, a theme that resonates with our results. Like Tkacová et al. (2022), we observed increased social media use for educational purposes during the COVID-19 pandemic, underscoring its potential as a learning tool. However, consistent with studies like Tkacová et al. (2022) and Adorjan and Ricciardelli (2021), we found that excessive social media use can detract from academic performance due to distractions and procrastination.

Our results highlight specific trends in social media usage and their correlations with academic performance metrics. For instance, students who use social media primarily for collaborative learning and accessing educational content tend to perform better academically. This supports the Social Learning Theory (SLT), which emphasises learning through observation and imitation, as students model positive academic behaviours they observe in their peers. Conversely, students who frequently use social media for entertainment or social interaction often exhibit lower academic performance, aligning with the Uses and Gratifications Theory (UGT). This suggests that the motivations behind social media use significantly influence outcomes, indicating that students seeking entertainment or social validation may experience more distractions and reduced study time.

**CONCLUSION AND RECOMMENDATIONS**

This study examined the correlation between students' utilisation of social media and their academic achievement. The results indicate significant engagement with social media platforms, with smartphones being the predominant means of access. Students generally use social media for socialising, while significant numbers also employ it for academic purposes. The research presents a multifaceted depiction of its impact on academic success. While some students perceive benefits such as improved comprehension through discussions and useful educational resources, others emphasise potential disadvantages such as heightened procrastination and decreased lecture attendance. Significantly, the study indicates the presence of a potential learning curve, as older students report fewer negative effects, possibly because they have established skills for effectively limiting their social media usage in an academic setting.

The study results propose several recommendations to enhance the positive impact of social media on academic performance while mitigating its drawbacks. Firstly, developing educational programmes to help students understand social media's potential benefits and drawbacks can equip them with effective strategies for managing their use. This initiative aims to raise awareness and provide students with the skills to balance social media engagement with academic responsibilities. Secondly, training faculty members to strategically integrate social media platforms into their teaching methods could significantly improve learning and engagement. Such training would focus on using social media as an educational tool while acknowledging and mitigating potential distractions. Lastly, exploring the development of mobile applications that facilitate focused academic use of social media platforms could prove beneficial. These apps would be designed to support academic activities and enhance productivity, ensuring that social media use contributes positively to students' educational experiences.

In order to address the potential detriments of social media on academic performance, it is recommended that specific educational programs and training methods be implemented. Digital literacy workshops can instruct students on using social media effectively for academic purposes, including identifying reliable sources, managing online time, and using social media tools for collaborative learning. Time management training sessions can help students balance social media use with academic responsibilities by teaching techniques such as creating study schedules and setting goals for social media usage. Furthermore, mindfulness and self-regulation programs can assist students in becoming more mindful of their social media habits and their impact on academic performance. Peer mentorship programs, where senior students share successful strategies with freshmen, can offer valuable practical insights. Also, integrating social media into the curriculum in a controlled manner can help students recognise its positive applications, such as using LinkedIn for networking or participating in academic discussions through class groups on Facebook.

Although this exploratory study illuminates the intricate relationship between social media usage and academic performance, certain limitations necessitate additional research. The sample size and survey design may not be representative of the entire student body, potentially limiting the generalizability of the findings. Additionally, the research relied on self-reported data, which can be susceptible to biases such as exaggeration or selective memory. To address these limitations, future research could explore a more diverse student population and utilise a combination of methods, including surveys, interviews, and time-management tracking tools, to gain a more comprehensive understanding of social media's impact on academic performance.

## ACKNOWLEDGEMENT

The authors appreciate the anonymous student participants used in this research. Also, Buntu Bewana's effort in data collection is greatly appreciated.

## REFERENCES

Acharya, Anita S, Anupam Prakash, Pikee Saxena, Aruna Nigam, and Anita Shankar Acharya. 2013. "Sampling: Why and How of It." *Indian Journal of Medical Specialities* 4 (2): 330–33. https://doi.org/10.7713/ijms.2013.0032.

Adorjan, Michael, and Rosemary Ricciardelli. 2021. "Smartphone and Social Media Addiction: Exploring the Perceptions and Experiences of Canadian Teenagers." *Canadian Review of Sociology/Revue Canadienne de Sociologie* 58 (1): 45–64. https://doi.org/10.1111/CARS.12319.

Al-Menayes, Jamal J. 2015. "Social Media Use, Engagement and Addiction as Predictors of Academic Performance." *International Journal of Psychological Studies* 7 (4): 86. https://doi.org/10.5539/IJPS.V7N4P86.

Amin, Zahid, Ahmad Mansoor, Syed Rabeet, Hussain And, and Faisal Hashmat. 2016. "Impact of Social Media of Student's Academic Performance." *International Journal of Business and Management Invention ISSN* 5:22–29. www.ijbmi.org.

Ansari, Jamal Abdul Nasir, and Nawab Ali Khan. 2020. "Exploring the Role of Social Media in Collaborative Learning the New Domain of Learning." *Smart Learning Environments* 7 (1): 1–16. https://doi.org/10.1186/S40561-020-00118-7/TABLES/5.

Bamigboye, Oluwatosin O., and A. Adelabu Olusesan. 2017. "An Analysis on Impact of Social Media for Learning in Eastern Cape Universities, South Africa." *International Journal of Educational Sciences* 17 (1–3): 69–75. https://doi.org/10.1080/09751122.2017.1305755.

Bandura, A. 2016. "The Power of Observational Learning through Social Modeling." In *Scientists Making a Difference: One Hundred Eminent Behavioral and Brain Scientists Talk about Their Most Important Contributions*, edited by R. J. Sternberg and S. T. Fiske D. J. Foss, 235–39. Cambridge University Press. https://psycnet.apa.org/record/2016-48867-050.

Bandura, A, and R. H Walters. 1977. *Social Learning Theory.* 1st ed. Prentice Hall: Englewood cliffs. https://psycnet.apa.org/record/1979-05015-000?utm_medium=email&utm_source=transaction.

Boyd, Danah M., and Nicole B. Ellison. 2007. "Social Network Sites: Definition, History, and Scholarship." *Journal of Computer-Mediated Communication* 13 (1): 210–30. https://doi.org/10.1111/J.1083-6101.2007.00393.X.

Gloria, Suita Allemina, and Surya Akbar. 2019. "THE IMPACT OF SOCIAL MEDIA USAGE TO ACADEMIC PERFORMANCE." *Jurnal Pendidikan Kedokteran Indonesia The Indonesian Journal of Medical Education* 8 (2): 68. https://doi.org/10.22146/JPKI.45497.

Goet, Joginder. 2022. "Impact of Social Media on Academic Performance of Students." *KIC International Journal of Social Science and Management* 1 (1): 35–42. https://doi.org/10.3126/KICIJSSM.V1I1.51100.

Griffioen, Nastasia, Hanneke Scholten, Anna Lichtwarck-Aschoff, Marieke van Rooij, and Isabela Granic. 2021. "Everyone Does It—Differently: A Window into Emerging Adults' Smartphone Use." *Humanities and Social Sciences Communications 2021 8:1* 8 (1): 1–11. https://doi.org/10.1057/s41599-021-00863-1.

Katz, E, J.G. Blumler, and M. Gurevitch. 1973. "Uses and Gratifications Research." *The Public Opinion Quarterly, JSTOR* 37 (4): 509–23. https://www.jstor.org/stable/2747854?casa_token=EE1sMGkxb_UAAAAA:BNSwrmkdNkT9UcbXCSijV2nxD7r3umoK7fUf97bhUJrDr2bbfGiADKjACTV4s7JWwXni2nLMrhu-IWA6zMAnq3y6NmZSHWh8RPOQWh9VkjiZtjkjG7M.

Khan, Muhammad Naeem, Muhammad Azeem Ashraf, Donald Seinen, Kashif Ullah Khan, and Rizwan Ahmed Laar. 2021. "Social Media for Knowledge Acquisition and Dissemination: The Impact of the COVID-19 Pandemic on Collaborative Learning Driven Social Media Adoption." *Frontiers in Psychology* 12 (May):648253. https://doi.org/10.3389/FPSYG.2021.648253/BIBTEX.

Lau, Wilfred W.F. 2017. "Effects of Social Media Usage and Social Media Multitasking on the Academic Performance of University Students." *Computers in Human Behavior* 68 (March):286–91. https://doi.org/10.1016/J.CHB.2016.11.043.

Lottering, R.A. 2020. "Using Social Media to Enhance Student Engagement and Quality." *South African Journal of Higher Education* 35 (4): 109–21. https://doi.org/10.20853/34-5-4271.

Lukose, Jose, Gardner Mwansa, Ricky Ngandu, and Olukayode Oki. 2023. "Investigating the Impact of Social Media Usage on the Mental Health of Young Adults in Buffalo City, South Africa." *International Journal of Social Science Research and Review* 6 (6): 303–14. https://doi.org/10.47814/IJSSRR.V6I6.1365.

Neelakandan, S., R. Annamalai, Sonia Jenifer Rayen, and J. Arunajsmine. 2020. "Social Media Networks Owing To Disruptions For Effective Learning." *Procedia Computer Science* 172 (January):145–51. https://doi.org/10.1016/J.PROCS.2020.05.022.

Odgers, Candice L., and Michaeline R. Jensen. 2020. "Adolescent Development and Growing Divides in the Digital Age." *Dialogues in Clinical Neuroscience* 22 (2): 143–49. https://doi.org/10.31887/DCNS.2020.22.2/CODGERS.

Odinokaya, Maria, Evgenia Tsimerman, Mohammad Akour, and Mamdouh Alenezi. 2022. "Higher Education Future in the Era of Digital Transformation." *Education Sciences 2022, Vol. 12, Page 784* 12 (11): 784. https://doi.org/10.3390/EDUCSCI12110784.

Rothkrantz, L. 2015. "How Social Media Facilitate Learning Communities and Peer Groups around MOOCS." *International Journal of Human Capital and Information Technology* 6 (1): 13. https://www.igi-global.com/article/how-social-media-facilitate-learning-communities-and-peer-groups-around-moocs/124215.

Rrustemi, Visar, Egzona Hasani, Gezim Jusufi, Dušan Mladenović, V Rrustemi, E Hasani, G Jusufi, D Mladenović, and Dušan Mladenovic. 2021. "Social Media in Use: A Uses and Gratifications Approach." *Management : Journal of Contemporary Management Issues* 26 (1): 201–17. https://doi.org/10.30924/MJCMI.26.1.12.

Ruggiero, Thomas E. 2000. "Uses and Gratifications Theory in the 21st Century." *Mass Communication & Society* 3 (1): 3–37. https://doi.org/10.1207/S15327825MCS0301_02.

Tkacová, Hedviga, Roman Králik, Miroslav Tvrdoň, Zita Jenisová, and José García Martin. 2022. "Credibility and Involvement of Social Media in Education—Recommendations for Mitigating the Negative Effects of the Pandemic among High School Students." *International Journal of Environmental Research and Public Health 2022, Vol. 19, Page 2767* 19 (5): 2767. https://doi.org/10.3390/IJERPH19052767.

Troussas, Christos, Akrivi Krouska, Efthimios Alepis, and Maria Virvou. 2020. "Intelligent and Adaptive Tutoring through a Social Network for Higher Education." *New Review of Hypermedia and Multimedia* 26 (3–4): 138–67. https://doi.org/10.1080/13614568.2021.1908436.